\documentclass[aps,prd,twocolumn,superscriptaddress,nofootinbib,floatfix]{revtex4-1}
\usepackage{graphicx,bm,dcolumn,color,epsfig,amsmath,amssymb,rotating,caption,float}
\usepackage[utf8]{inputenc}
\usepackage[T1]{fontenc}
\newcommand{\ba}{\begin{eqnarray}}
\newcommand{\ea}{\end{eqnarray}}
\newcommand{\non}{\nonumber}
\begin{document}

\title{Newtonian Potential and Geodesic Completeness in Infinite Derivative Gravity}
\author{Aindri{\'u} Conroy}
\author{James Edholm}
\affiliation{Consortium for Fundamental Physics, \\ Physics Avenue, Lancaster University, \\Lancaster, LA1 4YB, United Kingdom.}
\begin{abstract}
Recent study has shown that a non-singular oscillating potential -- a feature of Infinite Derivative Gravity (IDG) theories --
matches current experimental data better than the standard GR potential. In this work we show
that this non-singular oscillating potential can be given by a wider class of theories which allows the
defocusing of null rays, and therefore geodesic completeness. We consolidate the conditions whereby
null geodesic congruences may be made past-complete, via the Raychaudhuri Equation, with the
requirement of a non-singular Newtonian potential in an IDG theory.
In doing so, we examine a class of Newtonian potentials characterised by an additional degree of
freedom in the scalar propagator, which returns the familiar potential of General Relativity at large
distances.
\end{abstract}
\maketitle 

\section{Introduction}
In the century since Albert Einstein ushered in a new paradigm for modern physics by
formulating a gravitational theory that is described by the curvature of spacetime, General
Relativity (GR) has withstood numerous experimental tests~\cite{Will}.
%
It accurately describes our universe all the way down to short distances ($5.6 \times10^{-5}$m from a source \cite{Kapner:2006si})~\footnote{Dark energy notwithstanding} and has been found to be in agreement with 
experimental tests on gravitational redshift and the equivalence principle \cite{Clifton:2011jh}. More recently, the detection of 
gravitational waves lends  more weight to this already colossal theory~\cite{Abbott:2016blz}. However, that is not to say that GR 
does not have any shortcomings. At the classical level, the theory breaks down at short distances in its description of black holes, 
for instance. GR also cannot be made geodesically past-complete, when an appropriate energy condition is 
met, indicating the presence of an initial singularity in the theory~\cite{Vilenkin:2013tua,Kar:2008zz,Hawking1973,Vachaspati:1998dy}. 
In a geodesically-incomplete spacetime, causal geodesic congruences converge to a point in a finite `time' (affine parameter). In such a scenario a freely falling particle or photon will simply cease to exist in a finite `time', which suggests a serious physical malady in the theory \cite{Waldbook}. These shortcomings allow us to consider GR to be a first approximation to a broader theory. 

This notion of \emph{extending} GR, via additional curvature terms in the gravitational action, forms the basis for many modified theories of gravity. 
Significant examples include local theories such as $f(R)$-gravity, which replaces the curvature scalar in the Einstein-Hilbert action with an arbitrary function $f(R)$, 
or Stelle's 4th derivative gravity, which can be seen as a generalisation of the Gauss-Bonnet term in that it includes tensorial as well as scalar modifications to the Einstein-Hilbert action. 
However, these finite higher-derivative models still come hand-in-hand with a number of weaknesses. Stelle's theory, while perturbatively renormalizable \cite{Stelle}, suffers from the introduction of \emph{ghosts} 
- physical excitations, characterised by negative kinetic energy ~\cite{VanNieuwenhuizen:1973fi}. $f(R)$-gravity, in comparison, may avoid the introduction of ghosts \cite{Clifton:2011jh} but breaks down at short distances, meaning that the theory can not be said to be UV-complete \cite{Blas:2008uz}. In contrast, Infinite Derivative Gravity (IDG) offers a means of `completing' GR in the UV regime (short distances).
IDG is characterised by an action containing an infinite series of d'Alembertian operators ($\Box=g^{\mu\nu}{\nabla_\mu}{\nabla_\nu}$) 
acting on the curvature along with the mass scale of the theory, $M$.

The IDG action was first derived in~\cite{Biswas:2011ar}, where the form of the modified propagator was also calculated around a Minkowski background, see also \cite{Biswas:2013kla,Buoninfante:2016iuf}. The full non-linear equations of motion were computed in~\cite{Biswas:2013cha} and the boundary terms  found in~\cite{Teimouri:2016ulk}. The gravitational entropy for the IDG action was investigated around an (A)dS metric in~\cite{Conroy:2015nva}, while the form of the (A)dS propagator was given in \cite{Biswas:2016egy}. Constraints were put on the mass scale $M$ of IDG, either by using data on the tensor-scalar ratio and spectral tilt 
of the Cosmic Microwave Background~\cite{Edholm:2016seu}, or by looking at the deflection of light around the Sun~\cite{Feng:2017vqd}. 

In~\cite{Talaganis:2014ida,Talaganis:2015wva},
it was shown using a toy model of IDG that it is possible to curtail the divergences of 1-loop diagrams and show that 2-loop diagrams are finite, 
while in~\cite{Talaganis:2017tnr} 
the UV finiteness of IDG theories were investigated. Further work has focused on the Newtonian potential around a flat 
background~\cite{Biswas:2011ar,Edholm:2016hbt,Frolov:2015usa}, formulating the Hamiltonian of the IDG action~\cite{Mazumdar:2017kxr} 
and avoiding cosmological singularities both through using the ansatz $\Box R=c_1 R + c_2$~\cite{Biswas:2010zk,Biswas:2012bp} and  via the Raychaudhuri equation~\cite{Conroy:2016sac,Conroy:2017uds,Conroy:2014dja}. 

In this paper we investigate the Newtonian potential $\Phi(r)$, describing the gravitational field of a small static spherically symmetric test mass in a flat space background. In General Relativity, this diverges according to
$\Phi(r) \sim - 1/r$, becoming singular at the origin. IDG offers a means of resolving this divergence.

It was shown in \cite{Biswas:2013cha} that the equations of motion for IDG can be formulated in terms of two arbitrary functions of the d'Alembertian operator $a(\Box)$ and $c(\Box)$, which also characterise the modification to the gravitational propagator. In the case where these functions are equal, i.e. $a(\Box)=c(\Box)$, no additional degrees of freedom other than the massless graviton enter the system. Previous work \cite{Biswas:2011ar,Edholm:2016hbt} has shown that, in this case, a non-singular Newtonian potential can be derived, where the potential takes the form of an error function $\Phi\sim-\text{Erf}(Mr/2)/r$, for the simplest choice of $a(\Box)=e^{-\Box/M^2}$. If we generalise this further by taking $a(\Box)$ to be an exponential of a higher order polynomial, the potential is modified by an oscillating function at these higher orders.
Analysis by Perivolaropoulos~\cite{Perivolaropoulos:2016ucs, Kapner:2006si} has shown that this oscillating function provides a better fit to experimental data 
on the force of gravity at small distances than the standard GR theory. Although further analysis is needed, there are tantalising hints that modified gravity 
could provide the correct description of the strength of gravity at small distances. 

However, work on the avoidance of singularities in IDG theories~\cite{Biswas:2005qr,Conroy:2016sac,Conroy:2017uds,Conroy:2014dja} has shown that we require a departure from the simple choice of $a(\Box)=c(\Box)$ in order for causal geodesic congruences to be made past-complete. 
The aim of the present work is to consolidate the requirements of a non-singular theory of gravity, known as the \emph{defocusing} conditions, with the aforementioned converging Newtonian potential. To this end, we examine a wider class of Newtonian potentials, characterised by the condition $a(\Box)\neq c(\Box)$, in tandem with the defocusing conditions derived in \cite{Conroy:2016sac} around a Minkowski background. This would allow us to avoid singularities by permitting the defocusing of null rays in a theory with a well-defined Newtonian potential at short distances.

In Section \ref{sec:IDG}, we give an overview of Infinite Derivative Gravity (IDG), and show how the theory offers a means of rendering null rays geodesically-complete. In Section \ref{sec:newtonianpotential}, we derive the Newtonian potential for IDG, 
and specifically look at the case which allows defocusing. Finally in Section \ref{sec:plotting} we plot the Newtonian potential and interpret the results.
\section{Infinite Derivative Gravity}\label{sec:IDG}
As mentioned in the introduction, ghosts are physical excitations bearing negative kinetic energy. These excitations are represented by a negative residue in the gravitational propagator. When interactions in such a system take place, the vacuum decays into both positive and negative energy states. This is known as the Ostrogradsky instability~\cite{Woodard:2015zca}.

Previous attempts to resolve singularities by modifying gravity, such as Stelle's 4th derivative gravity~\cite{Stelle}, resulted in the introduction of ghosts, 
where the Hamiltonian of the theory was unbounded from below due 
to the Ostrogradsky instability \cite{Woodard:2015zca}. By adding an infinite number of derivatives to the theory this instability may be avoided through an appropriate choice of the functions $a(\Box)$ and $c(\Box)$. 
In \cite{Mazumdar:2017kxr} it was shown that the Hamiltonian of IDG is indeed bounded from below.

The IDG\ action, which is the most general, torsion-free and parity invariant action of gravity, that is quadratic in curvature was first derived in~\cite{Biswas:2011ar,Biswas:2005qr}
\ba
        S &=& \frac{1}{2} \int d^4 x \sqrt{-g} \bigg(M^2_p R + R F_1(\Box) R \non\\
        &&+ R^{\mu\nu} F_2(\Box) R_{\mu\nu} + C^{\mu\nu\lambda\sigma} F_3(\Box) C_{\mu\nu\lambda\sigma} \bigg),
\ea        
where $R$ is the Ricci curvature scalar, $R_{\mu\nu}$ is the Ricci tensor, $C_{\mu\nu\lambda\sigma}$ is the Weyl tensor and $M_p$ is the Planck mass. 
Each $F_i(\Box)=\sum_{n=0}^\infty f_{i_n} \Box^n/M^{2n}$ is a function of the d'Alembertian operator $\Box \equiv g^{\mu\nu} \nabla_\mu \nabla_\nu$.  
$M$ is the scale of modification of our theory and the $f_{i_n}$ are the dimensionless coefficients of the series. 

The equations of motion for IDG around a Minkowski background are given by~\cite{Conroy:2016sac}
\ba \label{eq:IDGeqationsofmotion}
       \kappa T_{\mu\nu} = a(\Box) R_{\mu\nu} - \frac{1}{2}\eta_{\mu\nu} c(\Box) R - \frac{1}{2}         f(\Box) \nabla_\mu \nabla_\nu R, 
\ea
where
we have defined\footnote{We denote the linearised curvatures around a Minkowski background as $R$, $R_{\mu\nu}$, $R_{\mu\nu\rho\sigma}$.}
\ba
        a(\Box) &=& 1 + M^{-2}_P \biggl( F_2(\Box) + 2 F_3(\Box) \biggr) \Box, \nonumber\\
        c(\Box) &=& 1 + M^{-2}_P \left( -4 F_1(\Box) - F_2(\Box) + \frac{2}{3} F_3(\Box)\right) \Box, \nonumber\\
        f(\Box) &=& M^{-2}_P \left( 4 F_1(\Box) + 2 F_2(\Box) + \frac{4}{3} F_3(\Box) \right), 
\ea
which abide by the constraint $a(\Box) - c(\Box) = f(\Box) \Box$. From (\ref{eq:IDGeqationsofmotion}) we can derive the propagator around a flat background~\cite{Biswas:2011ar,Biswas:2013kla}
\ba
        \Pi(k^2)= \frac{P^{(2)}}{k^2 ~a(-k^2)} - \frac{P^{(0)}_s}{k^2(a(-k^2)-3c(-k^2))}.
\ea 
The simplest choice is to set $a(\Box)=c(\Box)$ equal to the exponential of an entire function, which by definition does not have any roots,
 and therefore the propagator will not receive any additional degrees of freedom other than the massless graviton.
It was shown in~\cite{Buoninfante:2016iuf,Conroy:2017uds,Biswas:2005qr} that the scalar sector of the propagator can have at most one additional pole compared, without the introduction of ghost-like degrees of freedom.  
This allows us to write
\ba \label{eq:formofaminusthreec}
        (a(\Box)-3c(\Box)) R = 2\left( \Box/m^2 - 1\right)\bar{a}(\Box) R,
\ea
where $\bar{a}(\Box)$ is the exponential of an entire function and so $a(-k^2)-3c(-k^2)$ has a single pole in the scalar sector at $\Box\rightarrow -k^2 = m^2$, which produces an extra spin-0 particle of mass $m$. By expanding to first order in $\Box$,
we find that $m^2\equiv M^2_P/( 6 f_{1_0} -f_{2_0} - M^2_p/M^2)$~\cite{Conroy:2016sac}.
We know that $m^2>0$ so that the particle has real mass and therefore no tachyons are introduced.       

\subsection{Defocusing conditions}
The Raychaudhuri equation is a model-independent identity which relates the geometry of spacetime to the contribution of gravity via the curvature. It says that for a null tangent vector 
  $k_\mu$, satisfying $k^\mu k_\mu=0$,
the expansion parameter $\theta = \nabla_\mu k^\mu$ is described by\footnote{We have taken the simplest case of the Raychaudhuri
equation here by making two simplifications. Firstly, we take the congruence of null rays to be orthogonal to the hypersurface, so that the twist tensors vanish.
Secondly, the shear tensor gives a positive contribution to the right-hand side and so we can neglect it for our purposes.}
\ba
        \frac{d\theta}{d\lambda} + \frac{1}{2}\theta^2 \leq - R_{\mu\nu} k^\mu k^\nu,
\ea        
where $\lambda$ is the affine parameter~\cite{Waldbook}. In order to have an expansion parameter which is positive and increasing, and therefore allow the defocusing of the null rays, we require 
$R_{\mu\nu} k^\mu k^\nu<0$. In GR, using the Einstein equation, $G_{\mu\nu} = \kappa T_{\mu\nu}$, we cannot fulfil this condition because the Null Energy Condition (NEC) requires that $T_{\mu\nu} k^\mu k^\nu \geq 0$.
By the Hawking-Penrose Singularity Theorem, this inability to defocus will always lead to a singularity. To be precise, a spacetime cannot be null geodesically
complete in the past direction if $R_{\mu\nu} k^\mu k^\nu>0$~\cite{Hawking1973}.

However, in IDG this is not the case and we will see that it is possible to have defocusing.
By contracting the equations of motion (\ref{eq:IDGeqationsofmotion}) with $k^\mu k_\nu$, we can see that the contribution of gravity
to the Raychaudhuri equation is
\ba
        R_{\mu\nu} k^\mu k^\nu = \frac{1}{a(\Box)} \left[ \kappa T_{\mu\nu} k^\mu k^\nu 
        + \frac{k^\mu k^\nu}{2} f(\Box) \nabla_\mu \nabla_\nu R\right],~~
\ea    
which was studied in a cosmological setting~\cite{Conroy:2016sac}. 
When considering a static spherically symmetric perturbation around a flat background, then
the defocusing condition is\ba \label{eq:staticsphericallysymmraychaud}
        R_{\mu\nu} k^\mu k^\nu = \frac{1}{a(\Box)} \left[ \kappa T_{\mu\nu} k^\mu k^\nu 
        + \frac{(k^r)^2}{2} f(\Box) \Box R(r) \right]<0,~~~~~
\ea        
where $\Box R(r)=\frac{1}{r^2}\partial_r(r^2 \partial_r)R(r)$. Note that the function $a(\Box)$ acting on the curvature cannot be negative, as this would lead to the Weyl ghost. 
This is because $a(-k^2)$ is the modification to the spin-2 part of the propagator, 
so it must be positive to avoid negative residues~\cite{Biswas:2013kla}. Therefore if the NEC holds true, we arrive at the minimum defocusing condition  
\ba \label{eq:minimumdefocusingcondition}
        \frac{a(\Box)-c(\Box)}{a(\Box)} R(r) < 0.
\ea        

The first observation to make is that defocusing cannot occur in the case of $a(\Box)=c(\Box)$. 
However, if we allow an extra scalar propagating mode, from (\ref{eq:formofaminusthreec}) we can say that the relationship between $a(\Box)$ and $c(\Box)$ is given by~\cite{Conroy:2016sac}
\ba \label{eq:eqnforcintermsofaandatilde}
        c(\bar{\Box}) &=& \frac{a (\bar{\Box})}{3} \left[1 +2 \left( 1 -  \Box/m^2 \right) \tilde{a} (\bar{\Box})\right],
\ea 
where $\tilde{a}\equiv\frac{\bar{a}(\Box)}{a(\Box)}$ is an exponent of an entire function. Hence the propagator is given by~\cite{Conroy:2016sac}
\ba \label{eq:propagatorwithaneqc}
       \Pi(k^2) =\frac{1}{a(-k^2)} \left[ \frac{P^{(2)}}{k^2} + \frac{P_S^{(0)}}{2k^2(1+ k^2/m^2)\tilde{a}(-k^2)}\right],~~~~~~
\ea
while the minimum condition for null rays to defocus is (\ref{eq:minimumdefocusingcondition}) becomes
\ba \label{defocusingconditionwithaneqc}
        (1-\Box/m^2) \tilde{a}(\Box)R(r) > R(r).
\ea
\section{Newtonian potential}\label{sec:newtonianpotential}
When we take the metric generated by a small static spherically symmetric test mass added to a flat space background, following the method of~\cite{Biswas:2011ar,Conroy:2014eja},
\ba
        ds^2 = - \left(1+2\Phi(r)\right) dt^2 + \left(1-2\Psi(r)\right)\eta_{ij}dx^i dx^j,
\ea
this is akin to perturbing the flat space metric $\eta_{\mu\nu}$ as $g_{\mu\nu} = \eta_{\mu\nu} + h_{\mu\nu}$, where
\ba
           h_{00}=h^{00} = - 2 \Phi         \hspace{4mm} \text{and} \hspace{4mm}
        h_{ij} = h^{ij}=-2\Psi\eta_{ij}.
\ea
As a result, the scalar curvature and $00$ component of the Ricci curvature tensor around the flat Minkowski background are given by
\ba \label{eq:scalarcurvature}
        R = 4\Delta \Psi(r)  -2 \Delta \Phi(r), \quad \quad  R_{00} =  \Delta \Phi(r), 
\ea                 
where $\Delta \equiv \eta^{ij} \partial_i \partial_j$ is the Laplace operator. 
Then from the equations of motion (\ref{eq:IDGeqationsofmotion}), the trace and $00$ component equations of motion are \footnote{We have noted that the time derivatives of $R(r)$ and $R_{00}(r)$ vanish} 
\ba
       -\kappa \rho= \kappa T &=& \frac{1}{2}\left(a(\Box)  - 3 c(\Box) \right)R,\non\\
         \kappa\rho=\kappa T_{00} &=& a(\Box) R_{00} +\frac{1}{2} c(\Box) R. 
\ea   
Here we have taken the weak field approximation so that $\rho\gg p$, where $\rho$ is the density and $p$ is the pressure of the test mass. 
Therefore $ T= (-\rho+3p) \approx - \rho$ and $ T_{00} \approx \rho$. Combining this with
(\ref{eq:scalarcurvature}) leads to 
\ba \label{eq:psiintermsofphi}
        \Delta \Psi(r)  &=& -\frac{c(\Box)}{(a(\Box)  -2  c(\Box))}  \Delta \Phi(r),  \non\\
         \Delta \Phi(r)    &=& \frac{a(\Box)
         -2  c(\Box)}{a(\Box)\left(a(\Box)
         -3  c(\Box)\right)} \kappa \rho.
\ea
As our point source is of mass $\mu$, its density is approximated by a 3-dimensional Dirac-delta function:
$\rho = \mu\hspace{0.5mm}\delta^3(\boldsymbol{r})$.
Next we use a Fourier transform to calculate $\Phi(r)$ by using the same method as for calculating the Coulomb potential~\cite{schwartz2013,kiefer2012}. 
When the Dirac-delta function is Fourier transformed, it becomes~\cite{Biswas:2011ar} 
\ba
       \rho= \mu \hspace{0.5mm}\delta^3(\boldsymbol{r})=\mu \int \frac{d^3k}{(2\pi)^3} e^{i \boldsymbol{k} \cdot\boldsymbol{r}}.
\ea
As we go into momentum space, we take $\Box \to -k^2$, giving 
\ba \label{eq:phiofrintermsofaandc}
        \Phi(r)&=& - \frac{\kappa \mu}{(2\pi)^3} \int^\infty_{-\infty} d^3k \frac{a-2c}{a(a-3c)} \frac{e^{i \boldsymbol{k}\cdot \boldsymbol{r}}}{k^2}\nonumber\\  
        &=& - \frac{\kappa \mu}{4\pi^2 r} \int^\infty_{-\infty} dk \frac{(a-2c)}{a(a-3c)} \frac{\sin(kr)}{k}. 
\ea
Trivially, we can use (\ref{eq:psiintermsofphi}) to see that 
\ba \label{eq:Psiintermsofaandc}
        \Psi(r) = \frac{\kappa \mu}{4\pi^2
r} \int^\infty_{-\infty} dk \frac{c}{a(a-3c)} \frac{\sin(kr)}{k}. 
\ea
Note that if we set $a(-k^2)=c(-k^2)$, 
\ba \label{eq:newtonianpotentialforaequalsccase}
        \Phi(r)= \Psi(r)= - \frac{\kappa \mu}{4\pi^2
r} \int^\infty_{-\infty} dk \frac{1}{2a} \frac{\sin(kr)}{k}. 
\ea
This case was discussed in~\cite{Biswas:2011ar,Edholm:2016hbt}. For the simplest case, where $a(\Box)=e^{-\Box/M^2}$, it results in the $1/r$ 
fall of the potential seen in GR being modified by the error function, i.e. $\Phi(r) \sim \frac{\text{Erf}(r)}{r}$. 
For large $r$, $\text{Erf}(r)\approx 1$ but at short distances, $\text{Erf}(r)\sim r$, so $\Phi(r)\sim 1$. 
Therefore the $1/r$ behaviour is retained at large distances but at short distances, the modification means that
the potential simply tails off to a constant and the potential is no longer singular.

\subsection{Potential for IDG with defocusing}          
In this paper we will extend the calculation of the IDG potential to the case where $a\neq c$, which allows us to avoid singularities.
We describe the relationship between $a(\Box)$ and $c(\Box)$ using (\ref{eq:eqnforcintermsofaandatilde}), and by inserting (\ref{eq:eqnforcintermsofaandatilde})
into (\ref{eq:phiofrintermsofaandc}) we find that
\ba \label{eq:phiintermsofatildeandm}
        \Phi(r)  &=& - \frac{\kappa \mu}{24\pi^2
r} \non\\ \times&&\int^\infty_{-\infty} dk \left[ 4-\frac{m^2}{\tilde{a}(-k^2)(m^2+k^2)}\right] \frac{\sin(kr)}{k~a(-k^2)}.~~~~
\ea   
        The calculation we need to perform is 
\ba \label{eq:generalfofrforaneqc}
        f(r)=\int^\infty_{-\infty} dk \left[ 4-\frac{m^2}{\tilde{a}(-k^2)(m^2+k^2)}\right] \frac{\sin(kr)}{k~a(-k^2)},~~~~ 
\ea 

where $\Phi(r)\sim - \frac{f(r)}{r}$. We can write $\tilde{a}(-k^2)= e^{\tau(-k^2)}$ and $a(-k^2)= e^{\gamma(-k^2)}$, and this results in
\ba \label{eq:finalcalculation}
        f(r)= \int^\infty_{-\infty} dk \left[4 -\frac{m^2 e^{-\tau(-k^2)}}{m^2+k^2}\right]
\frac{e^{-\gamma(-k^2)}\sin(kr)}{k}.~~~~~
\ea
This is our main result - we have shown that we can have defocusing as well as a non-singular Newtonian potential. 
This potential  returns to the GR value in the infrared limit, i.e for large values of $r$.\footnote{Note that in the limit 
$M\to \infty$ we return to a non-local theory, which is fourth-order gravity~\cite{Gao:2012fd}. }

\subsection{Conditions on $a(-k^2)$ and $\tilde{a}(-k^2)$}
Next we investigate the conditions that must be placed on $a(-k^2)$ and $\tilde{a}(-k^2)$, and therefore what we can say about 
$\tau(-k^2)$ and $\gamma(-k^2)$.  

First we look at the spin-0 part of the propagator. From (\ref{eq:propagatorwithaneqc}), we have
\ba \label{eq:propagatorintermsofegammaandetau}
       \Pi(k^2) = \frac{1}{e^{\gamma(-k^2)}} \left[ \frac{P^{(2)}}{k^2} + \frac{P^{(0)}}{2k^2(1+
k^2/m^2)e^{\tau(-k^2)}}\right],~~~~~
\ea
where $P^{(2)}$ is the spin-2 projection operator and $P^{(0)}$ is the spin-0 projection operator. 
We require that in the UV, the propagator is exponentially suppressed, allowing us the possibility of curtailing divergences \cite{Talaganis:2014ida}. 
From the spin-0 sector of \eqref{eq:propagatorintermsofegammaandetau}, we observe the condition $\gamma(-k^2)+\tau(-k^2)>0$. 
Fortunately, this condition ensures that the integral (\ref{eq:finalcalculation}) converges.
Note that a priori, it is possible that there are coefficients in front of the exponentials
$e^{\gamma(-k^2)}$ and $e^{\tau(-k^2)}$, but by requiring that we return to the GR propagator in the infrared limit $k\ll {m, M}$, we are obliged to set the coefficients to 1. 
Also, by requiring the spin-2 part of the propagator (\ref{eq:propagatorintermsofegammaandetau}) to be exponentially suppressed we have $\gamma(-k^2)>0$.

From looking at (\ref{defocusingconditionwithaneqc}) in momentum space, the defocusing condition imposes
a constraint: $\tau(-k^2)\geq0$. 
Thus, we have the functions $a(-k^2) = e^{\gamma(-k^2)}$ and $\tilde{a}(-k^2)=e^{\tau(-k^2)}$ where $\gamma(-k^2)$ is positive function and $\tau(-k^2)$ is a 
non-negative function.
In fact, the result of these constraints is that the integral (\ref{eq:finalcalculation}) converges.

We therefore have two free mass scales $M$ and $m$ and two unspecified functions $\tau(-k^2)$ and $\gamma(-k^2)$. 
We will henceforth choose both the mass scales to be the Planck mass $M_P$ and choose the simplest possible version of $\tau(-k^2)$, which is $\tau$~=~0 . 
This choice means that all of the freedom in the model is tied up in the function $\gamma(-k^2)$, which we will choose to be the monomial $\gamma(k^2)= (Ck^2/M^2_P)^D$.
We therefore have only two free parameters, $C$ and $D$.

\section{Plotting the results} \label{sec:plotting}
\begin{figure}[h]
\includegraphics[width=9cm]{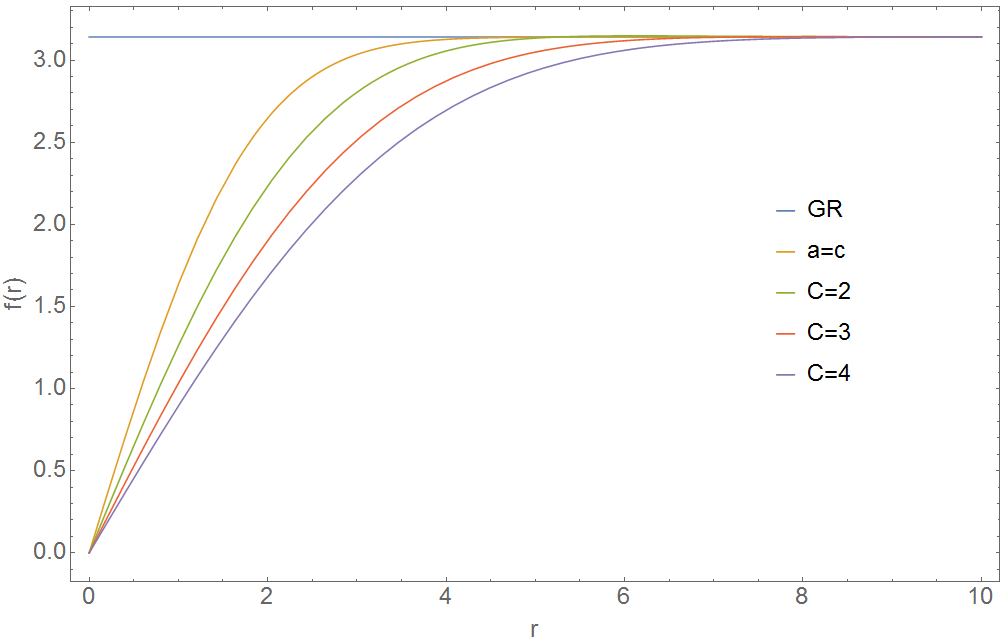} 
\caption{We plot $f(r)$ vs $r$ for different $C$ and $D=1$, where $\gamma(k^2)=(Ck^2/M^2_P)^D$. See (\ref{eq:defocusingboxfr}). 
For illustrative purposes, we have taken $M_p=1$m$^{-1}$. We compare our results to the $a(\Box)=c(\Box)$ case seen in (\ref{eq:newtonianpotentialforaequalsccase}). 
Here, we see that as $C$ increases,  the effect of IDG can be seen increasingly further away from the origin.}
\label{Fig1}
\end{figure}
\begin{figure}[h]
\includegraphics[width=9cm]{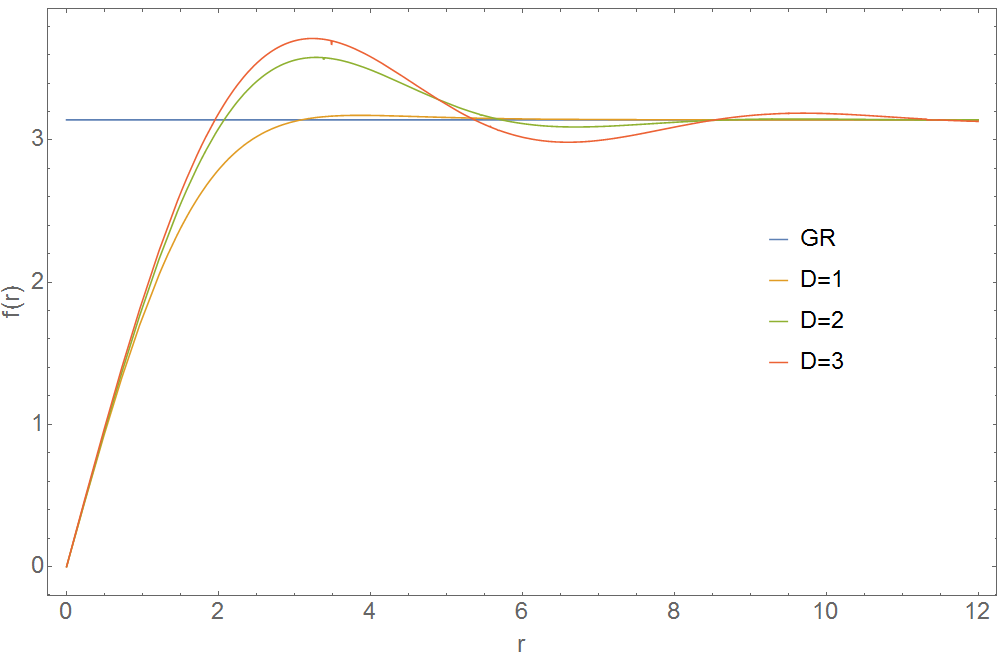}
\caption{We plot $f(r)$ vs $r$ for different $D$ and $C=1$ where $\gamma(k^2)=(Ck^2/M^2_P)^D$. See (\ref{eq:defocusingboxfr}). 
The plot shows that for $D>1$ the potential oscillates. As the value of $D$ increases, so too does the magnitude of these oscillations. 
For illustrative purposes, we have taken $M_p=1$m$^{-1}$ and again we find that our results reduce to that of GR at large distances.}
\label{Fig2}
\end{figure}
\begin{figure}[h]
\includegraphics[width=9cm]{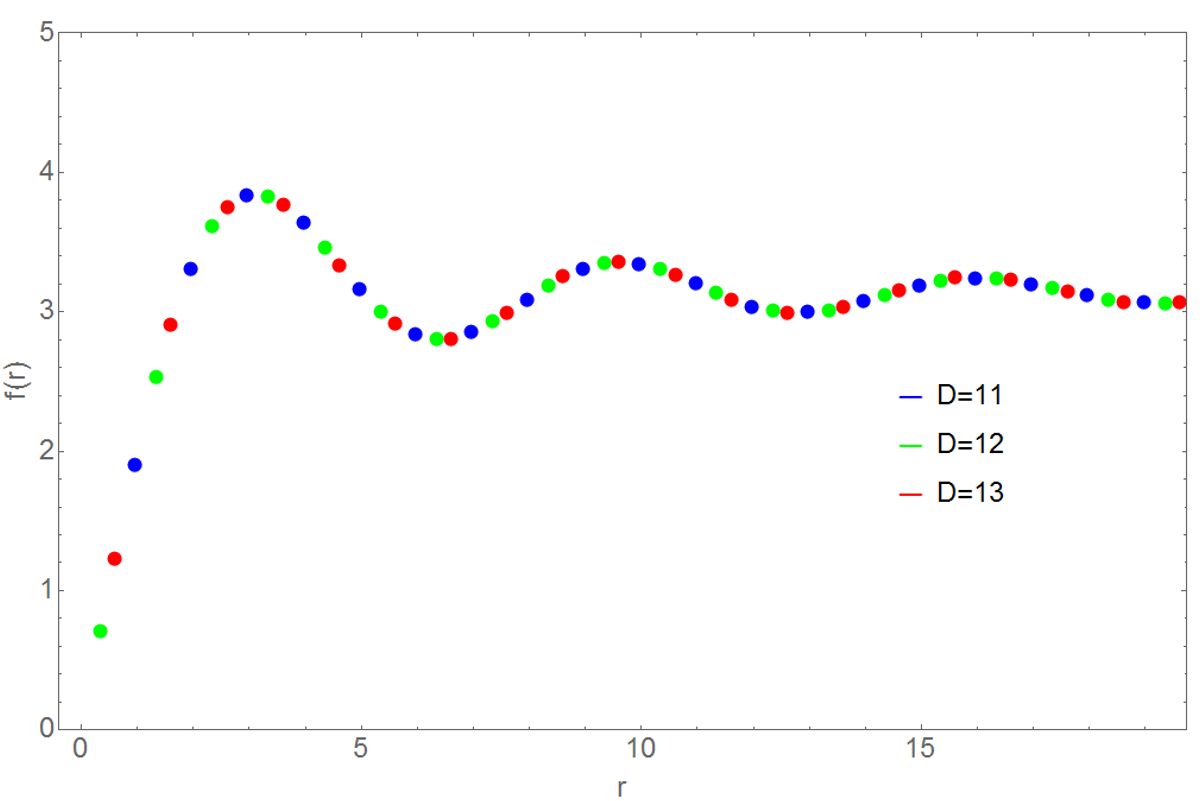}
\caption{We plot $f(r)$ vs $r$ for different $D$ greater than 10 with $C=1$, where $\gamma(k^2)=(Ck^2/M^2_P)^D$. See (\ref{eq:defocusingboxfr}).
We can see that for $D>10$, 
increasing $D$ does not affect the potential. 
For illustrative purposes, we have taken $M_p=1$m$^{-1}$. 
 We can parameterise these curves as $\alpha_1r$ for $r<1$ and $\alpha_2 \cos(\theta r +\theta_0)/r$ for $r>1$, 
where $\alpha_1$, $\alpha_2$, $\theta$ and $\theta_0$ are constants, as in~\cite{Perivolaropoulos:2016ucs}.}
\label{Fig3}
\end{figure}
\subsection{Choosing a form for $a(-k^2)$ and $\tilde{a}(-k^2)$}
For the simplest choice $\tau(-k^2)=0$ and $\gamma(k^2)=(Ck^2/M_P^2)^D$, this gives\footnote{Note that for large $r$, 
then $\Psi(r)= \Phi(r)$, i.e. the Eddington parameter $-\Phi/\Psi$ is equal to one, as expected from experimental bounds using
data from the Cassini probe~\cite{Deser:2007jk}.}  
\ba \label{eq:defocusingboxfr}
        f(r)= \int^\infty_{-\infty} dk \left[4 -\frac{M_P^2 }{M_P^2+k^2}\right]
\frac{e^{-(Ck^2/M_P^2)^D}\sin(kr)}{k}.~~~~~~
\ea

In Fig.~\ref{Fig1}, we take $D=1$ and plot $f(r)$ for different choices of $C$ using (\ref{eq:defocusingboxfr}).  
Clearly increasing the value of $C$ moves the point at which the non-locality kicks in further away from the origin.
Next in Fig.~\ref{Fig2} we look at how the potential varies for the same $C$ but with different values of $D$, which is the power of $k^2$ in (\ref{eq:defocusingboxfr}). 
As we increase the value of $D$, our potential begins to oscillate, as was found in \cite{Edholm:2016hbt}.
These oscillations grow in size as $D$ increases until about $D=10$. 

This is because
\ba \label{eq:rectanglefunction}
        \lim_{D\to \infty} e^{-(Ck^2/M_P^2)^D}=\mbox{rect}(Ck^2/M_P^2),
\ea
where $\mbox{rect}(x)$ is the rectangle function,
which is defined by $\mbox{rect}(x)=1$ for $|x|<1$ and $\mbox{rect}(x)=0$ for $|x|>1$. For $D>10$, (\ref{eq:rectanglefunction}) is a very good approximation and
so increasing the value of $D$ does not change the potential.

Our next task is to investigate the choice of large powers of $k^2$ which has been shown to fit recent experimental data~\cite{Perivolaropoulos:2016ucs} at small distances for the $a(\Box)=c(\Box)$ case. 
We will see whether we can still obtain oscillating solutions with conditions necessary to realise defocusing. 
In Fig.~\ref{Fig3} we plot (\ref{eq:defocusingboxfr}) for different choices of $D>10$ and note that it can still be parameterised accurately as
\ba
        f(r) =
        \begin{cases}
                 \alpha_1 r  &\text{for } 0<r<1, \nonumber\\
                 1 + \alpha_2 \frac{\cos(\theta r +\theta_0)}{r} &\text{for }1<r.
        \end{cases}
\ea

In other words, the oscillating solution which was hinted at by experimental data can be produced by a modified gravity solution which also allows geodesic completeness. 
\section{Conclusion}
We have found the Newtonian potential for a wider class of Infinite Derivative Gravity (IDG) theories than were 
previously investigated and analysed various cases of the theory. Analysis of data from experimental tests has hinted 
that, at small distances, an oscillating non-local potential provides the best fit to experimental data. We have shown that an IDG theory constrained to allow the defocusing of null rays, and therefore geodesic completeness, 
still produces a non-singular potential which returns to the standard GR result at large distances. This result can still be parameterised as the oscillating function which provides a good fit to the data. 

By allowing defocusing, it is necessary to introduce extra parameters into the model, although we can reduce the freedom in these parameters 
by making appropriate choices about the mass scales and the form of the functions. 
Our results can be tested experimentally, which will allow us to put constraints on our parameters. 
Future research could look at moving away from the minimal model and reintroducing the choice of parameters which we reduced, or finding the potential around a de Sitter background rather than a flat space background. 

\section{Acknowledgements}
The authors would like to thank David Burton and Anupam Mazumdar for their invaluable help in preparing this paper.


\end{document}